\documentclass[12pt]{article}                                  
\usepackage{epsfig,rotating}
\usepackage{cite}
\usepackage{axodraw}

\newcommand{\dz}{{\mbox{d}}z}

\newcommand{\beq}{\begin{equation}}
\newcommand{\eeq}{\end{equation}  }

\def\q{\mathrm{q}}
\def\go{\mathrm{g}}
\def\g{\mathrm{\gamma}}

\def\d{\delta}

\def\E{\mbox{e}^+\mbox{e}^-}

\def\O{\Omega}
\def\L{\mathrm{M}}
\def\S{\mathrm{T}}
\def\T{\Theta}
\def\E{\mbox{e}^+\mbox{e}^-}
\def\cd{\mathcal{D}}
\def\cn{\mathcal{N}}

\begin{document}
\clearpage
\pagestyle{empty}
\setcounter{footnote}{0}\setcounter{page}{0}%
\thispagestyle{empty}\pagestyle{plain}\pagenumbering{arabic}%

\hfill ANL-HEP-PR-98-21
 
\hfill February, 1998

\vspace{1.0cm}

\begin{center}

\vskip 0.8in plus 2in
%%%%%%%%%%%%%%%%%%%%%%%%%%%%%%%%%%%%%%%%%%%%%%%%%%%%%%%%%%%%%%%
%                   1997                              %
%%%%%%%%%%%%%%%%%%%%%%%%%%%%%%%%%%%%%%%%%%%%%%%%%%%%%%%%%%%%%%%
{\Large\bf Scale-Invariant Dynamical Fluctuations \\ 
in Jet Physics\\[-1cm]} 
%%%%%%%%%%%%%%%%%%%%%%%%%%%%%%%%%%%%%%%%%%%%%%%%%%%%%%%%%%%%%%%

\vspace{2.0cm}

\renewcommand{\thefootnote}{\fnsymbol{footnote}}

{\large S.V.Chekanov 
\footnote[1]{On leave from
Institute of Physics,  AS of Belarus,
Skaryna av.70, Minsk 220072, Belarus.}
}

\vspace{1.0cm}

Argonne National Laboratory, 
9700 S.Cass Avenue, 
Argonne, IL 60439
USA

\vspace{1.5cm}
PACS numbers: 12.38 Aw, 12.40 Ee, 13.85 Hd

\bigskip
Short title: Scale-Invariant Fluctuations

\vspace{1.5cm}

\begin{abstract}
An interpretation of  
scale-invariant multiplicity fluctuations  inside
hadronic jets is presented. It
is based on the branching mechanism
with the angular ordering of 
soft partons  in sequential branchings. 
A relationship with fractal distributions 
is demonstrated.
The model takes into account the 
finiteness of the number of particles
produced in jets (finite energy) 
and leads to a good description of the
multifractal fluctuations
observed in  $\E$ processes.
\end{abstract}

\end{center}
\newpage
\setcounter{page}{1}

%%%%%%%%%%%%%%%%%%%%%%%%%%%%%%%%%%%%%%%%%%%%%%%%%%%%%%%%%%%%%%%%%%
\section{Introduction}
\label{sec:int}
%%%%%%%%%%%%%%%%%%%%%%%%%%%%%%%%%%%%%%%%%%%%%%%%%%%%%%%%%%%%%%%%%%

In the case of $\E$ 
annihilation processes the asymptotic collinear and 
infrared contributions to gluon cross sections can be 
described in Double Leading Log Approximation 
(DLLA) by a Markov process (see \cite{bas} for a review). 
This semi-classical description takes into
account soft gluon interference effects on the basis of the
angular ordering prescription when the parton emission is described by
successive branchings and the available phase space is reduced
to ever smaller angular regions (color coherence effects). 
The corresponding QCD master equation  is an integral one and 
is based on  Dokshitzer-Gribov-Lipatov-Altarelli-Parisi 
energy-distribution kernels.

In the framework of this description, progress has been made in 
obtaining  angular scale-invariant\footnote{The 
scale-invariance means that a dynamical characteristic $X(l)$
of correlations/fluctuations at a given resolution $l$ has the  property
$X(\lambda l)=\lambda^{-L}X(l)$  with   a constant $L$  characterizing
dynamics of a multiparticle system.}
correlations between partons \cite{an1} 
(see \cite{an2} for a review). 
This approach,  by conception,  is
{\em a correlation} one, based on the method
of characteristic functionals. Hence, to derive 
directly measurable quantities such as normalized factorial
moments or factorial cumulants, one needs to
perform an integration of the correlation functions over the restricted
phase-space region under study. This is possible only after the use of many 
approximations and
by identifying the phase-space regions which
give the leading contributions \cite{an2}.

Apart from this problem, there are also more basic 
questions which restrict the direct comparison of the 
QCD correlation  approach with experimental data.
Firstly, the perturbative QCD calculations deal with an {\em asymptotic}
behavior of the multiparton correlations valid only for very high
energies. In an idealized jet, therefore,   
finite parton multiplicities in small phase-space bins and
energy-momentum conservation effects are
systematically  ignored \cite{an1}. This  is one of the
most important reasons leading to disagreements between
the analytical predictions and $\E$ data \cite{exper1,experl3}. 
Secondly, the increase of the coupling constant for very small
phase-space regions sets a limit for the validity of  perturbative QCD.
Thirdly, non-perturbative 
effects such as hadronization, resonance
decays and Bose-Einstein correlations complicate the comparison 
of theoretical many-particle inclusive
densities with the data even at  LEP1 energies 
\cite{exper1,experl3}.

In this paper, therefore,  we propose a new way to study 
the correlations in terms of fluctuations 
in the multihadronic  systems produced
in  high-energy processes. Being
based on a {\em fluctuation} approach to intermittency phenomenon
(see recent reviews \cite{int11,int1} on the subject of 
intermittency), the model  {\em a priori}
takes into account the finiteness of the number 
of particles in a single event (finite energies).   
In order to describe the local multiplicity
fluctuations, we adopted the differential Markov equation for
parton branching, which has been  
used to describe global multiplicity
fluctuations in high-energy physics in \cite{mar1,ge,mar11,mar2}
(see also references in \cite{int1}).

One of the key ideas of this  approach is that, in contrast
to a full phase space, a  Markov branching 
process inside a small phase-space window of size $\d$ 
can be characterized by a  probability $P_n(t,\d )$ 
of detecting $n$ particles, 
in which a dependence on an evolution parameter 
$t$ can be factorized from a  phase-space $\d$-dependence
(see Sect.~\ref{local}).
As a consequence of this assumption, 
the  scale-invariant fluctuations  experimentally 
observed inside jets \cite{bpexp} may be 
considered as a result of fractal phase-space distribution 
for each particle emitted in successive Markov branchings 
(Sect.~\ref{locm}).
Such an idea  ultimately leads to the possibility of taking  into
account an inhomogeneity of the parton correlations inside a jet
and a fairly good quantitative agreement with the 
$\E$-annihilation data \cite{bpexp} 
and the JETSET 7.4 PS model \cite{jetset}
(Sect.~\ref{mpred}).

%%%%%%%%%%%%%%%%%%%%%%%%%%%%%%%%%%%%%%%%%%%%%%%
\section{Statistical treatment of branchings}
\label{local}
%%%%%%%%%%%%%%%%%%%%%%%%%%%%%%%%%%%%%%%%%%%%%

%%%%%%%%%%%%%%%%%%%%%%%%%%%%%
\subsection{Global equation}
\label{glo}
%%%%%%%%%%%%%%%%%%%%%%%%%%%%

At high energies, gluons dominate the parton-parton cross 
section due to the large color factor and the infrared singularity. 
This means that a good 
high-energy approximation  should  consider  gluon branching only.
For generality, however, we shall consider both  gluons and
quarks treating them as partons.

Let $t$ be the evolution parameter of the parton  branching process.
The $t$ can be related to the parton virtuality $Q$ and can be 
defined in the usual way \cite{mar1,ge,mar11,mar2}. 
However, hereafter we 
shall never refer to the explicit form of this parameter and shall
regard it as  representing the extent of branching or just time.
We assume that the branching process starts with $t=0$
and continues until some $t_{\mbox{max}}$ determined by a QCD cut-off $Q_0$.
The initial condition for the probability distribution $P_n(t)$
of having $n$ particles radiated by the initial one  is
\beq
P_{n=0}(t=0)=1, \qquad P_{n\ne 0}(t=0)=0.
\label{2}
\eeq

In the following we shall see that, under the assumptions to be made below, 
the structure of local fluctuations 
depends  neither  on  the particular definition 
of the evolution parameter, nor on the initial conditions.
The purpose of  the introduction of (\ref{2}) is only
to give an illustration of the notion of a typical initiation  of the
cascade and its further evolution.

A probabilistic scheme \cite{bas} of the perturbative parton shower is based
on classical picture of the Markov chains of independent
parton splittings. Each elementary parton 
decay depends on just the nearest ``forefather''.
Let us define $W_1\mbox{d}t$ as the probability of 
branching $a\to b+c$ during a small range of $t$, $\mbox{d}t$,
according to one of the following decays: $\go\to\go\go$, $\q\to\q\go$,
and $\go\to\bar{\q}\q$.
The infinitesimal  probability $W_1$ in the leading log
picture can be written as 
\beq 
W_1=\sum_{a,b}\int_0^1\dz \frac{\alpha_{s}}{2\pi}P_{a\to bc}(z),
\label{qw0}
\eeq
where $\alpha_{s}$ is the strong coupling constant and $P_{a\to bc}(z)$ are 
the Dokshitzer-Gribov-Lipatov-Altarelli-Parisi
energy-distribution kernels. 
The sum runs over all allowed  parton branchings. 
For our simplified model, we will consider
the case with $\alpha_{s}=const$, so that 
$W_1$ is a (divergent) constant which does not depend on $t$. 
 
If  there are $n$ partons, the probability of the parton emission increases.
Let $W_{n}\mbox{d}\,t$ be the probability that the parton system 
with multiplicity $n$ radiates a new parton during   
the infinitesimal interval ($t$, $t+\Delta t$).
Generally, $W_{n}$ depends  on  the parton multiplicity $n$.
This  can be taken into account as  
\beq
W_n=w(n)W_1,  \qquad  w(1)=1,
\label{qw1}
\eeq 
where $w(n)$ is a  function of $n$ reflecting  
an increase of the parton radiation. 
Then the Markov pure birth evolution equation 
for the multiplicity distribution  $P_n(t)$ of having 
$n$ partons at time $t$ is well-known \cite{mapr}:
\beq
\frac{\partial P_n (t)}{\partial t} =
W_{n-1}P_{n-1}(t) -  W_nP_n(t).
\label{2a}
\eeq
The solution of this equation  is a global multiplicity
distribution $P_n(t)$. Since the equation contains ingredients of
perturbative QCD, an essential point is to regularize $W_1$ and
consider the branching
evolution up to $t_{\mbox{max}}$ determined by the QCD cut-off  $Q_0$.
In order to compare the
$P_n(t_{\mbox{max}})$  with the data,
one usually resorts to  the local parton-hadron duality hypothesis
which states that $n$ for partons is proportional to the
$n$ for observed hadrons.

The  differential  equation  (\ref{2a}) with
constant ($t$-independent) vertex probabilities  $W_n$ 
has been analyzed  in \cite{mar1,ge,mar11,mar2}.
One of the most popular solutions is a 
negative binomial distribution which was derived 
in the leading log picture for gluons in quark jet \cite{ge}. 
Deviations from this distribution observed 
in $\E$ annihilation data
are usually connected with the shoulder structure and 
a  quasi-oscillatory behavior of $H_q$ moments seen at $Z^0$ peak. 
Recently, however, the negative binomial distribution
has been  reestablished again: 
In was shown that the full-phase-space 
multiplicity distribution for  $\E$ annihilation data  can be well reproduced 
by a weighted superposition of two negative 
binomial distributions \cite{re1,re2},
associated to  two-  and
multi-jet events or the contributions  from  
$b\bar{b}$ and light flavored events.    

For the full phase space, there is no physical reason to define
$W_n$ in momentum space: The global distribution is
momentum independent. However, to obtain 
various momentum characteristics of particle spectra 
(such as the multiplicity of partons above a fixed momentum), 
a more complex integro-differential equations should be
analyzed \cite{mar1,bas}.  
Below we will discuss another way to
include a momentum dependence using a statistical projection of
equation (\ref{2a}) into momentum phase-space domains. 

%%%%%%%%%%%%%%%%%%%%%%%%%%%%%%%%%%%%%%%%%%%%%%%%%%%%%%
\subsection{Local equation}
\label{loc}
%%%%%%%%%%%%%%%%%%%%%%%%%%%%%%%%%%%%%%%%%%%%%%%%%%%%%%

Obviously, if one 
counts  only the particles produced within a certain small 
range of phase space, not all  particles can  be detected in it. 
Let $\g_n(\d )$ be the probability of observing one particle 
in a  phase-space domain of  size $\d$ if this particle
belongs  to the parton system of multiplicity $n+1\ge 1$ 
in the full phase space.  
We put
\beq
0\le \g_n(\d ) \le 1,
\label{3}
\eeq
so that
\beq
\g_n(\d =0)=0, \qquad \g_n(\d =\Omega )=1, 
\label{4}
\eeq
where $\Omega$ is the size of full phase space ($\d\le\Omega$)
which  can be defined in 3-momentum phase space or, say,
in rapidity, $p_{\mbox{t}}$ or azimuthal angle. 

For  a  phase-space element of size $\d$, 
if the system is in state $n$ at time $t$, the probability of the transition
$n\to n+1$ in the interval $(t, t+\Delta t)$ is
$$
\g_n(\d )W_n\Delta t + o(\Delta t,\d ), 
$$
where, as before, $W_n$ describes the emission of one particle 
into the full phase-space $\Omega$ and 
the factor $\g_n(\d )$ describes the
probability of hitting $\d$ by this  particle. 
The factorization property of the infinitesimal
probability $\g_n(\d )W_n$ is an essential assumption  
used to simplify the  structure of parton evolution.
We also  assume that the probability $P_n(t+\Delta t, \d)$  of having 
$n$ particles inside $\d$ at $t+\Delta t$ is fully
determined by $P_n(t, \d)$ and  $P_{n-1}(t, \d)$ in 
the same $\d$. In fact, for a particular (``angular'') choice of 
phase space, this is consistent with the coherent branching with
angular ordering, since
the contribution of particles from phase-space regions
outside of $\d$ is considered to be very small 
(see the discussion below). On the basis of
these assumptions, one can write 
$$
P_n(t+\Delta t,\d )=\g_{n-1}(\d )W_{n-1}P_{n-1}(t,\d)\Delta t +
(1-\g_n(\d)W_n\Delta t)P_n(t,\d ) + o(\Delta t,\d ),
$$
where the second term is due to probability conservation.
Then the  corresponding Markov equation for  the {\em local}
multiplicity distribution $P_n(t,\d )$ is 
\beq
\frac{\partial P_n(t,\d )}{\partial t}=\g_{n-1}(\d )W_{n-1}P_{n-1}(t,\d )
- \g_n(\d )\,W_n P_n(t,\d ).
\label{5}
\eeq
As we see,  from the point of view
of an observer counting particles in $\delta$, the restriction
of the phase-space domain  looks as  an effective suppression
of the birth rate  $W_n$. (c.f.(\ref{2a})). 
Note also that, in contrast to (\ref{2a}), equation (\ref{5})
contains a momentum dependence via $\g_n(\d )$.

It is  necessary to note  that  condition (\ref{3}) comes from
a probabilistic interpretation of $\g_n(\d )$.
Generally, as $W_n$, this quantity
can be larger than unity.  
However, if this is the case,  we can carry out
the following transition: $\g_n (\d )\to Z\g_n (\d )$, where
$Z$ is a constant, so that the
condition (\ref{4}) for  $Z\g_n (\d )$ can hold.
As we shall see below, this  regularization
does not change the structure of observable fluctuations 
derived from (\ref{5}). 

Clearly, a possible non-linear nature of  equation (\ref{5}) 
renders its explicit solution very difficult.
It can be solved in a  straightforward manner   only 
for some particular forms of the 
vertex probabilities $W_{n}$  and  $\g_n(\d )$.

%%%%%%%%%%%%%%%%%%%%%%%%%%%%%%%%%%%%%%%%%%%%%%%%%%%%%%%%%%%%%%%
\subsection{Phase-space property in the factorization scheme}
\label{sec:equ}
%%%%%%%%%%%%%%%%%%%%%%%%%%%%%%%%%%%%%%%%%%%%%%%%%%%%%%%%%%%%%%%

We will be interested  in a  general solution of (\ref{5})
with respect to the possible behavior of the probability
$P_n(t, \d )$ as a function of $\g_n (\d )$. 

For $n=0$, the solution can be easily obtained  
\beq
P_0(t,\d )=\exp\left(-\g_0(\d ) \int W_0\mbox{d} t\right).
\eeq
This exponential form of  $P_0$ is similar to  the Sudakov
form factor. In contrast to the full phase space, 
the integral contains  the suppression
factor $\g_0(\d )$ taking into account the fact that 
a particle can be emitted 
outside of  the small phase-space interval.

The form of $P_n(t,\d )$ for $n\ge 0$ cannot be obtained 
without knowing the form  of $W_n$ and $\g_n(\d )$. However, 
a  phase-space structure of such a solution can be deduced
in a general case. Since the basic  idea of this approach
is to factorize the phase-space and $t$-dependent component, let 
us look for the solution of (\ref{5}) in the
form
\beq
P_n(t ,\d )=f_n(t) p_n(\d ) + o(t,\d ), \qquad n\ge 1,
\label{s1}
\eeq
where $f_n(t)$ is a $\d$-independent and $p_n(\d )$ is $t$-independent
well integrable functions. We assume that (\ref{s1}) 
has a sense for any $t$ 
at  a sufficiently small $\d$. 
 
Using (\ref{s1}), (\ref{5}) can be rewritten as 
\beq
\frac{p_1(\d )}{P_{0}(t,\d )}=\g_0(\d )b_1,
\label{s19}
\eeq
\beq
\frac{p_n(\d )}{p_{n-1}(\d )}=
\g_{n-1}(\d )\, b_n, \qquad n\ge 2, 
\label{s20}
\eeq
\beq
b_n=\frac{W_{n-1}f_{n-1}(t)}{f_n^{'}(t) +f_n(t)W_n\g_n(\d )},
\qquad f_0(t)=1.
\label{s20e}
\eeq
Since we are looking for a solution at small $\d$, $\g_n(\d )$
has a  small value. Therefore, $b_n$ can be approximated by 
the $\d$-independent constant,
\beq
b_n\simeq \frac{W_{n-1}f_{n-1}(t)}{f_n^{'}(t)}.
\label{s20g}
\eeq 
Further, the assumption (\ref{s1}) holds  only if
$b_n$ is independent of $t$ for $n\ge 2$.
For a given $W_{n}$, (\ref{s20g})  can be solved with respect to the
form of $f_n(t)$.  
However,  the $\d$-dependence of $P_n(\d ,t)$ has 
already been obtained. It reads
\beq
\frac{P_n(t,\d )}{P_{n-1}(t,\d )}
\simeq \g_{n-1}(\d )\, b_n\, \frac{f_n(t)}{f_{n-1}(t)}\simeq
\g_{n-1}(\d )W_{n-1}\frac{f_n(t)}{f_{n}^{'}(t)},
\qquad n\ge 1.
\label{s20h}
\eeq
Let us remind that this relation is assumed to be possible only if 
$\d$  is  small. In this case, the solution for $P_n(\d ,t)$
may be factorized as in (\ref{s1}) 
(see an example in subsection~\ref{sub:f}). 
 
Of course, to study  the distribution $P_n(t,\d )$  
as a function of $\d$  by means of 
factorial moments or cumulants might technically be  a very difficult task. 
However, having in mind the bunching-parameter method 
\cite{bp1,bp2,bp3}, this distribution
can easily be  analyzed. Bunching parameters (BPs) $\eta_q(t,\d )$ 
are defined as
\beq
\eta_q(t,\d )=\frac{q}{q-1}
\frac{P_q(t,\d )P_{q-2}(t,\d )}{P_{q-1}^2(t,\d )}.
\label{mot1}
\eeq
They  measure the deviation of the  multiplicity
distribution $P_n(t,\d )$ from a Poisson one  for which
the BPs are equal to unity. Generally, in the case of no
dynamical phase-space correlations, $\eta_q(t,\d )$ are 
$\d$-independent. 

The BP of an arbitrary order $q$ for (\ref{s20h})  can be written as
\beq
\eta_q(t,\d )=\eta_q(t)\, \eta_q(\d ),
\label{18c}
\eeq
\beq
\eta_q(t)=\frac{w(q-1)}{w(q-2)}\frac{f_qf_{q-1}^{'}}{f_q^{'}f_{q-1}}, 
\label{18xc}
\eeq
where $\eta_q(\d )$  depends only on the phase-space interval, 
\beq
\eta_q(\d )=\frac{q}{q-1}\frac{\g_{q-1}(\d )}{\g_{q-2}(\d )}.
\label{18cc}
\eeq

As we see, the structure of $\eta_q(t,\d )$  is quite remarkable.
It contains a $t-$dependent function $\eta_q(t)$ 
constructed from unknown 
$w(n)$ and $f_n(t)$, so that  equation (\ref{5}) itself  
can have  strong non-linear property. However, since we study
the fluctuations at ever smaller $\d$, this function   is 
unrelevant: 
The property of the local fluctuations  
is fully determined by the ratio $\g_{q-1}(\d )/\g_{q-2}(\d )$.

Note that while the original equation (\ref{5}) is  constructed from 
the divergent constants  $W_n=w(n)W_1$, the final result
for the BPs  does not contain them directly, since
$W_1$ cancels  in (\ref{18xc}).
However, (\ref{5})  contains them indirectly via $f_n(t)$.
We can handle this problem since
the regularization 
procedure $\g_n(\d )\to Z\,\g_n(\d )$ discussed
in subsection~\ref{loc}
does not change the BPs (\ref{18xc}) and, 
hence, the observable fluctuations.
According to this, one can always redefine 
$\g_n(\d )$ as $\g_n(\d )\to W^{-1}_1\g_n(\d )$,
so that $W_1$ cancels already in (\ref{5}). 

%%%%%%%%%%%%%%%%%%%%%%%%%%%%%%%%%%%%%%%
\subsection{Markov birth-death process}
%%%%%%%%%%%%%%%%%%%%%%%%%%%%%%%%%%%%%%%

The same phase-space behavior (\ref{18cc}) of the BPs 
can be obtained from a stationary 
Markov birth-death evolution equation.
For small $\d$, this process has to be characterized
by the birth rate $\g_n(\d ) W_{n}^{+}$ and 
the death rate $W_{n}^{-}$ due to 
the fusion (absorption) processes such as
$\go\go\to\go$,  $\q\go\to\q$ and
$\bar{\q}\q\to\go$. These effects  are  not important
for the full phase-space. However, for small $\d$,   
the values of $\g_n(\d ) W_{n}^{+}$ and  $W_{n}^{-}$
can be comparable. The local equation reads
$$
\frac{\partial P_n(t,\d)}{\partial t}=\g_{n-1}(\d )W_{n-1}^+P_{n-1}(t,\d)
+ W_{n+1}^-P_{n+1}(t,\d) -
\left[\g_n(\d )\,W_n^+ + W_n^-\right]P_n(t,\d).
$$
Assuming that for very small $\d$ the process is a stationary,
$\partial P_n/\partial t \sim  0$, one can derive  
(see details in \cite{mam0})
\beq
\frac{P_n(\d )}{P_{n-1}(\d )}=
\frac{W_{n-1}^+}{W_n^-}\,\g_{n-1}(\d ),
\label{14}
\eeq
which is similar to (\ref{s20h}). Hence, BPs have the same
form as (\ref{18c}), with the phase-space dependence 
as in (\ref{18cc}). The only difference is 
that $\eta_q(t)$ in (\ref{18xc}) does not depend on $t$
and has the form:
\beq
\eta_q=\left(\frac{W_{q-1}^+}
{W_{q-2}^+}\right)\left(\frac{W_{q-1}^-}
{W_{q}^-}\right).
\label{18d}
\eeq 

Note that the  stationary (equilibrium)  regime is a strong
assumption. It cannot be applied to the full phase space.
For local phase-space domains, the physical situation is
somewhat different:  
Each emitted parton increases the
phase space for further emissions and  
the total phase space is expanded with increasing $t$.
However, if one counts the particles inside a
selected small phase-space
window, one  may assume that 
there is  a little change in the density 
of partons inside $\d$ with
increasing $t$ and, hence, 
$P_n(t,\d)$ does not depend strongly on $t$.
This assumption can be verified  experimentally
by observing $t$-independence of the BPs.

%%%%%%%%%%%%%%%%%%%%%%%%%%%%%%%%%%%%%%%
\subsection{Fully independent emission}
\label{sub:f}
%%%%%%%%%%%%%%%%%%%%%%%%%%%%%%%%%%%%%%%

A simple example of the approach discussed 
above provides a fully independent particle  emission. 
For this  we should use the following assumptions:

\medskip

1) $W_n$ in (\ref{5}) does not depend on $n$, i.e. $w(n)=1$, 
$W_n=W_1$;

2) $\g_n(\d )$ does not depend on $n$, $\g_n(\d )=\g(\d )$. 

\medskip

Under these conditions, equation (\ref{5}) can be easily solved.
The solution is a Poisson distribution, 
\beq
P_n(\d )=a^n\exp(-a)/n!, \qquad a=W_1\, t\, \g(\d ).
\label{pois}
\eeq
The behavior of this distribution at small $\d$ can be factorized
as in (\ref{s1}),
$$
P_n(\d )\simeq (W_1\, t)^n \g^n(\d )/n! + o(\g^n(\d )),  
$$
so that the corresponding BPs are unity.
Note that for (\ref{pois}) this is true not
only locally ($\d\to 0)$, but also for any $\d$.  
For a uniform phase-space distribution,
$\g(\d)$ is simply equal to $\d/\Delta$.

Generally, an independent {\em phase-space} 
particle production can be characterized by any $W_n$ with
$\g_n(\d )=\g(\d )$. In this case the BPs are $\d$-independent
constants.

%%%%%%%%%%%%%%%%%%%%%%%%%%%%%%%%%%%%%%%%
\section{Local fluctuations in the model}
\label{locm}
%%%%%%%%%%%%%%%%%%%%%%%%%%%%%%%%%%%%%%%%%
%%%%%%%%%%%%%%%%%%%%%%%%%%%%%%%%%%%%%%%%%%%%%%%%%%
\subsection{Statistically averaged picture of a jet}
%%%%%%%%%%%%%%%%%%%%%%%%%%%%%%%%%%%%%%%%%%%%%%%%%%

To study the phase-space dependence of fluctuations,  the next step is
to understand a  possible behavior of $\g_n(\d )$ in (\ref{18cc}).

\begin{figure}
\begin{center}
\begin{sideways}
\begin{sideways}
\begin{sideways}
\mbox{\epsfig{file=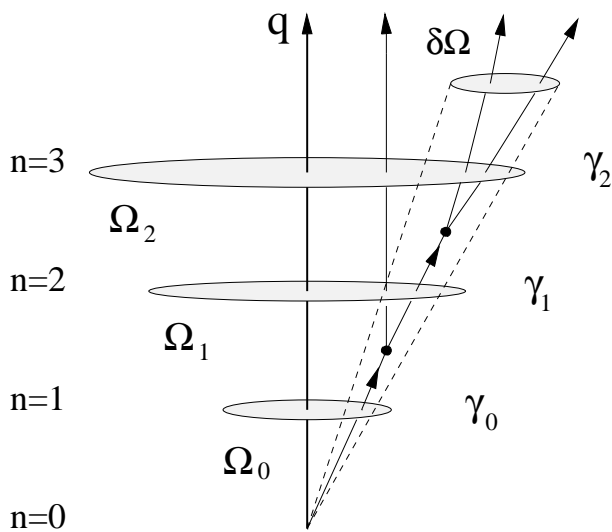,height=8.0cm}}
\end{sideways}
\end{sideways}
\end{sideways}
\caption[jet]
{\it 
A schematic  representation of the phase-space
structure of  branching inside jet. It makes use the
angular ordering prescription: The structure of the cascade
inside $\d\O$ is determined by the ``history'' of this cascade
inside the same $\d\O$. Infinitesimal  probabilities $W_n$ (not shown) 
control the structure of the cascade for  full phase space $\O_n$.
Local infinitesimal  probabilities $\g_nW_n$ determine
the structure of  cascade inside $\d\O$.}
\label{jet}
\end{center}
\end{figure}

We shall start our consideration with a
simple two-dimensional model of a jet in angular intervals.
Let us consider the first parton emitted at
some angle with respect to the initial quark. Since we are
interested in a picture  averaged  over all events,
let $\O_0$
be the maximum possible size of solid angle,
so that the first parton  always has  an  angle
inside the cone $\O_0$ (see Fig.~\ref{jet}).
After its emission, the first parton radiates the next one at
some angle with respect to its own direction of flight.
Generally, we assume that there is recoil effect
and the first parton can change its direction
after this  radiation.
In this case,  the  solid angular window available for both
partons becomes larger  and is  equal to $\O_1>\O_0$.
The second parton then splits into two new partons at
$\O_2$ and so on.
One can  further simplify the model taking into
the account  angular ordering when available phase space is
reduced for successive branchings. In this case
$\O_0\simeq\O_1\simeq\O_2\simeq\ldots\O_n$.

Let us tern to a more detailed description in one dimension. 
First, let us define $\T$ as 
the polar angle between the 
directions of motion of the emitted  and the parent
parton. The single-particle distribution 
$\rho (\T )$ of the gluon bremsstrahlung 
can be approximated \cite{an1,an2} at small $\T$ by 
\beq
\rho (\T )=  C(Q_0, \alpha_s) \T^{-1}, 
\label{pk1}
\eeq
integrating the overall distribution over the azimuthal angle around the
quark direction and momentum dependence. The $\T$-independent constant
$C(Q_0, \alpha_s)$  contains  a transverse momentum cut-off $Q_0$ and 
$\alpha_s$ which is treated here as a  constant. 
The probability $\g_0(\d\T )$  of finding the gluon inside the small interval 
$(\T_0-\d\T, \T_0)$ near a jet  opening angle $\T_0$ is
\beq
\g_0(\d\T )\propto  \int_{\T_0-\d\T}^{\T_0}\rho (\T )\mbox{d}\T
\sim  \d\T^{\cd_0},
\quad \cd_0=1
\label{pk2}
\eeq
for $\d\T\to 0$.
Note that this result does not depend essentially 
on the details of the density  $\rho (\T )$, 
since it  has no  singularity near  $\T_0$.
We did not specify a coefficient of proportionality
between $\g_0(\d\T )$ and $\d\T^{\cd_0}$: As we have seen
before, the phase-space dependence of  
the fluctuations does not depend on it.

Now let us consider the behavior of $\g_1(\T )$ for the second parton. 
Since we are interested in the probability of emission
of this parton into $(\T_0-\d\T, \T_0)$  under the condition that 
the first parton is inside the same interval, 
there is a larger probability of hitting
this interval by the second parton because of the 
collinear singularity. Now the major problems  in
the calculating $\g_1(\T )$ are: 1)
An ambiguity in the position of the first parton inside $\d\T$;
2) Singularity of $\rho (\T )$ near $\T\sim 0$ gives a dominant
contribution. This leads to a very inhomogeneous
phase-space distribution near $\T_0$; 
3) Requirement of the angular ordering. 

Due to the reasons quoted above, the calculation
of $\g_n(\T )$ for $n>1$ is  even more difficult.
We shall make no attempts to calculate $\g_n(\T )$.  
In a general case, for small $\d$, we assume
\beq
\g_n(\d\T )\propto \d\T^{\cd_n}, \qquad n\geq  1,
\label{fr11}
\eeq
where $\cd_n$ are $\d\T$-independent 
constants controlling  the collinear singularities together
with  the angular ordering restrictions of the phase space 
available for particles on  $(n+1)$th multiplicity stage.
The latter effect decreases the available phase space
for the  next soft offspring partons that would
increase the probability of  detecting  them inside $\d\T$.
We assume,
\beq
\cd_0\geq \cd_1\geq \cd_2\geq \ldots \geq \cd_\infty.
\label{fr11x}
\eeq 
In subsection~\ref{sec:frac} 
we shall give an interpretation of the behavior 
(\ref{fr11}) and (\ref{fr11x})
in terms of fractal distributions.  
Then we shall see that the  behavior of 
$\g_n(\d\T )$ for small $\d\T$  is 
the {\em only simplest choice}  which allows
to describe experimental   data.
In Sect.~\ref{mpred} we shall  proceed with 
the physical interpretation of these quantities.

\medskip

There are a number of special cases of interest:

\medskip

1) {\em Monofractal fluctuations}

This  case corresponds to the situation when  the  phase-space 
distributions  for all cascade  stages
(except the initial one)  have the same non-uniformity  
characterized by 
$\cd_1$, i.e.,
\beq
\g_0(\d\T )\propto \d\T^{\cd_0}, 
\qquad  \g_{n>0}(\d\T )\propto \d\T^{\cd_1}. 
\label{mon}
\eeq

Making use of (\ref{18c}), the BPs are
\beq
\eta_2(\d\T )\propto 
\d\T^{\cd_1 - \cd_0},
\qquad
\eta_{q>2}(\d\T )=const.
\label{mon2}
\eeq
Hence,  we obtain  the monofractal
behavior with $d_2=\cd_0 - \cd_1$ \cite{bp1,bp2}.

For cascade branchings, such a situation
can be considered as a highly unrealistic since it totally
disregards  that daughter partons have ever
larger probability to be emitted inside $\d\T$ because of
the correlations. Therefore, the monofractal type of
intermittency possibly observed for some nucleus-nucleus reactions
may mainly be attributed to other dynamical mechanisms \cite{sec},
rather than to actual cascade processes with angular ordering.

\medskip

2) {\em Multifractal fluctuations}

If particles on each cascade stage   
are distributed differently,
then the cascade stage with the multiplicity 
$n+1$  should be characterized
by its own $\cd_n$, i.e.,
\beq 
\g_{n\ge 0}(\d\T )\propto \d\T^{\cd_n}.  
\label{mul}
\eeq
The corresponding BPs are
\beq
\eta_q(\d\T )\propto 
\d\T^{-\alpha_q}, 
\qquad \alpha_q=\cd_{q-2} - \cd_{q-1}.
\label{mul2}
\eeq
An  inverse relation for $\cd_n$ reads
\beq
\cd_n = \cd_0 -\sum_{i=2}^{n+1}\alpha_i .
\label{mul2a}
\eeq
According to \cite{bp1,bp2}, one has
a multifractal behavior. An example  of such a behavior will
be given in subsection~\ref{sec:ex}.

%%%%%%%%%%%%%%%%%%%%%%%%%%%%%%%%%%%%%%%%%%%%%%%%%%%%%%%
\subsection{Connection with fractals}
\label{sec:frac}
%%%%%%%%%%%%%%%%%%%%%%%%%%%%%%%%%%%%%%%%%%%%%%%%%%%%%%%

The simplicity of the model allows  a natural connection of it
with fractals. In this subsection we shall see that $\cd_n$
introduced in (\ref{fr11}) are  nothing but fractal dimensions.

First, let us  remind 
a standard definition of a fractal distribution.
Let us assume that there is a large number 
$\cn_{\mbox{tot}}$ of particles
distributed  over a phase space with  the
topological (Euclidean) integer
dimension $D$ ($D=1,2,3$). 
Let $\cn (\epsilon )$ be the number of particles
counted inside the phase-space domain with a  linear
size $\epsilon$. The number $\cn (\epsilon )$
and $\epsilon$ are related as  
\beq
\cn (\epsilon )\propto  \epsilon^{\cd}, \qquad \epsilon\to 0,
\label{dd11}
\eeq
where $\cd$ is a fractal dimension,
corresponding  to the so-called  box-counting
(or mass, cluster $etc.$) dimension \cite{ffr}. 
If the distribution is
extremely inhomogeneous, $\cd$ has a 
non-integer value ($\cd <D$).
If particles were uniformly distributed
over the phase space, $\cd$ is integer ($\cd=D$). 
Therefore, $\cd$ is  a very economical 
way to describe the extent of non-uniformity of a distribution
near a given small phase-space region. 
 
It is easy to see that (\ref{dd11}) also characterizes 
the probability $p$ of observing one particle inside $\epsilon$:
This  probability  
is determined by the ratio of the number $N(\epsilon )$ 
of events of observing  a  particle  
inside $\epsilon$ to the 
total number $N_{\mbox{tot}}$ of events. 
Assuming that only one particle can be emitted 
in each event, one has
\beq
p\equiv\frac{N(\epsilon )}{N_{\mbox{tot}}}=
\frac{\cn (\epsilon )}{\cn_{\mbox{tot}}}
\propto  \epsilon^{\cd}, \qquad \epsilon\to 0.
\label{dd1}
\eeq

Now let us tern to the model. 
In fact, the $\g_n(\d\T)$ has the same meaning as the $p$ defined 
in (\ref{dd1}).
The index $n$ in $\g_n(\d\T)$ simply specifies  the cascade
stage $n$ with the total $n+1$ particles, so that 
$\cd_n$ stands the  fractal dimension of the phase-space distribution
of a single particle on each cascade stage.
Then  (\ref{pk2}) describes  a uniform particle distribution 
near $\T_0$ (no collinear singularity!). 
For the second particle, 
there is no such a uniformity any more: The collinear singularity
of the emission of the second particle is near $\T_0$ and this
leads to a very inhomogeneous distribution in this region, 
so that $\g_1(\d\T )\propto \d\T^{\cd_1}$, where $\cd_1$ is
a  fractal dimension of this distribution ($\cd_1<\cd_0 =1$). 
For the next  emissions,
the distribution should be even more inhomogeneous since
parent particles are already non-uniformly
distributed due to the collinear singularities and the angular ordering. 
Finally this leads to the condition $\cd_n \geq \cd_{n+1}$ 
guessed in (\ref{fr11x}). 

The  $\cd_n$ are  the  usual fractal dimensions. 
However, after many cascade steps with different $\cd_n$, 
one obtains  a  multifractal behavior (\ref{mul2}) of the BPs.
For  a monofractality (\ref{mon2}),   
the phase-space distribution for each particle in the cascade 
has to be  characterized by a single fractal dimension for all $n$, 
$\cd_0\ne \cd=\cd_1=\cd_2=\ldots$.

%%%%%%%%%%%%%%%%%%%%%%%%%%%%%%%%%%%%%%%%%%%%%%%%%%%
\subsection{Connection with factorial-moment method}
%%%%%%%%%%%%%%%%%%%%%%%%%%%%%%%%%%%%%%%%%%%%%%%%%%%%

A widely used means to study  local fluctuations is
based on the calculation of
the normalized factorial moments \cite{rcas}:
\beq
F_{q}(\d\T )=\frac{\langle n(n-1)\ldots
(n-q+1)\rangle }{\langle n\rangle ^q},
\label{nfm1}
\eeq
where $n$ is the number of particles inside a
restricted phase-space interval $\d\T$,
$\langle\ldots\rangle$
is the average over all events.
For non-statistical fluctuations,  $F_{q}(\d\T )$ depend on
the size of the phase-space interval $\d\T$  as
$F_{q}(\d\T )\sim\d\T^{-\phi_{q}}$, where
$\phi_{q}$ are intermittency indices.

If the size of  phase space is asymptotically small,
then the
following approximate relation between the $F_q(\d\T )$
and the BPs  holds \cite{bp1,bp2}:
\begin{equation}
F_{q}(\d\T )\simeq\prod_{n=2}^{q}\eta_{n}^{q+1-n}(\d\T ).
\label{nfm2}
\end{equation}
From (\ref{nfm2}) and (\ref{mul2}), one has
\beq
F_q(\d\T )\propto\d\T^{-\phi_q}, \qquad
\phi_q = \sum_{n=2}^{q}(q-n+1)\alpha_n,
\label{fmop}
\eeq
or, taking into account the expression for $\alpha_n$, 
\beq
\phi_q = \sum_{n=2}^{q}(q-n+1)\,(\cd_{n-2} - \cd_{n-1}).
\label{fmop9}
\eeq
The case of no dynamical correlation corresponds to
$\phi_q=0$. From (\ref{fmop9}), it follows that the only
possibility for this case is the condition
\beq
\cd_0 =1=\cd_1= \cd_2 = \ldots .
\label{fmop2}
\eeq
i.e., the next emitted partons are distributed
over available  phase space  purely randomly (uniformly).

The model allows a simple way to connect the 
R\'{e}nyi fractal dimension (see details in \cite{int1}) 
for factorial moments with the usual fractal dimensions $\cd_q$
in our model. The R\'{e}nyi fractal dimension $D_q$ is defined
via $\phi_q$, 
\beq
D_q=D-\frac{\phi_q}{q-1}.
\label{pp1}
\eeq
From (\ref{fmop9}) one has,
\beq
D_q=\cd_1 - \sum^{q}_{n=3}\frac{q-n+1}{q-1}
(\cd_{n-2} - \cd_{n-1}),
\label{pp2}
\eeq
where we take into account that
the topological dimension $D$ is equal to $\cd_0$. 
From here one can again see that the monofractality ($D_q=const$)
is possible only if $\cd_{n-1} = \cd_{n}$, for $n>1$.
A variation of $D_q$ with $q$ for the multifractal case
can be due to  $\cd_{n-1} \ne  \cd_{n}$. 

In fact, the information about the fractal dimensions
$\cd_n$ can be extracted from the study of both
$D_q$ (for factorial moments) or $\alpha_q$ (for bunching parameters). 
However, the study of the BPs is the most direct way
to obtain the information on $\cd_n$:

\medskip

1) In contrast to the BPs,  
the power-like behavior of the
normalized factorial moments  
holds only approximately 
for one dimensional variables
because of a saturation effect for  small
rapidity intervals  (see \cite{int11,int1,bpexp}).

\medskip  
2) The BP $\eta_q$ of  order $q$  is a differential tool,
resolving only the difference $\cd_{q-2}-\cd_{q-1}$ between
the fractal dimensions $\cd_n$ (see (\ref{mul2})). 
In contrast, the normalized
factorial moment $F_q$ of  order $q$  is an ``integral'' tool, 
which is sensitive to   
to all $\cd_n$ with  $n-1<q$. Because of the factor in the sum (\ref{pp2}),
the contribution from  $\cd_n$  at small $n$ is the largest.
Hence, small changes in the behavior of $\cd_n$ for 
large $n$ may be hidden
due to contributions from $\cd_n$ for small $n$.

%%%%%%%%%%%%%%%%%%%%%%%%%%%%%%%%%%%%%%%%%%%%%%%%%%%%%%%
\subsection{Experimental data}
\label{sec:ex}
%%%%%%%%%%%%%%%%%%%%%%%%%%%%%%%%%%%%%%%%%%%%%%%%%%%%%%%

The multifractal behavior (\ref{mul2}) of BPs is
characteristic for many different  reactions 
\cite{bp1}. For example, for rapidity variable
with respect to the trust axis,  BPs depend
on the size of rapidity interval $\d y$ as   
\beq
\eta_q(\d y)=\beta_{q}^{'} \, \d y^{-\alpha_q}, \qquad q\ge 2,
\label{mot4}
\eeq
where $\beta_{q}^{'}$ and $\alpha_q$
are positive constants.
This can be considered as an evidence that local
fluctuations have a scale-invariant structure, $\eta_q(\lambda\d y)=
\lambda^{-\alpha_q}\eta_q(\d y )$, i.e. the 
behavior is invariant under change of scale.

Usually, the power law (\ref{mot4}) is represented in 
terms of the number $M=Y/\d y$
of   bins of size $\d y$ covering a full phase-space volume $Y$,
so that (\ref{mot4}) becomes
\beq
\eta_q(M)=\beta_{q}^{'}\, M^{\alpha_q}.
\label{mot5}
\eeq
Taking the logarithm from  both sides, the power law can be
written as the linear expression
\beq
\ln \eta_q(M)=\alpha_q\ln M + \beta_q, \qquad \beta_q=\ln \beta_{q}'.
\label{mot6}
\eeq

\begin{figure}
\begin{center}

\vspace{-2.5cm}
\mbox{\epsfig{file=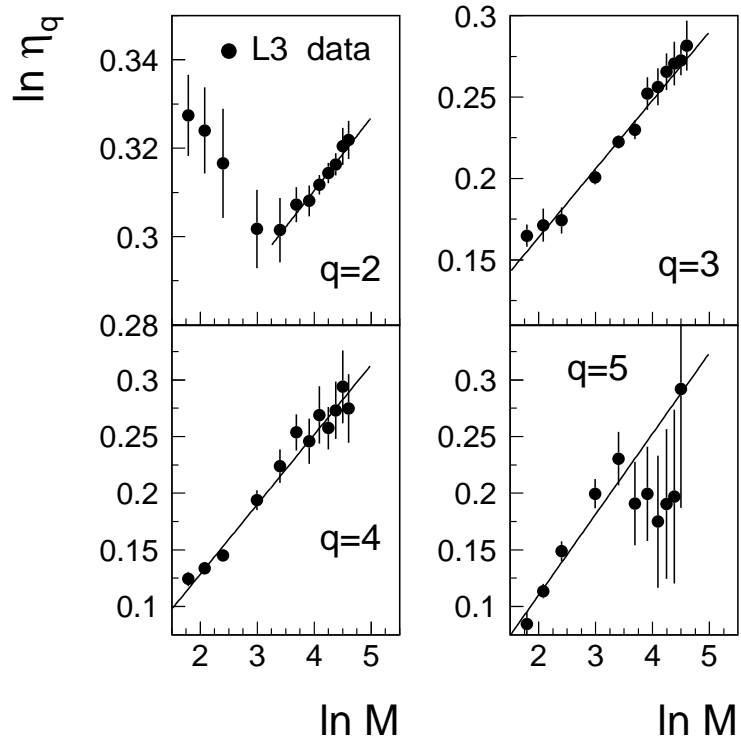,width=10.0cm}}
\caption[bpmark]
{\it BPs for  rapidity defined with respect
to the thrust axis for $\E$ processes.
Here $M=Y/\d y$, where $Y$ is the size of full rapidity interval,
$\d y$ is the restricted rapidity interval.
The data are  reproduced from \cite{bpexp}.
The lines represent the fit by (\ref{mot6}) 
with the parameters presented in
Table~\ref{tab1}.}
\label{bpmark}
\end{center}
\end{figure}

For $\E$ annihilations, such a behavior
has been  observed for  rapidity defined with respect to the thrust
axis (see Fig.~\ref{bpmark} and \cite{bpexp,bp1,bp2}). 
That the $\alpha_q$ are not zero and vary with $q$ is a direct indication
that the fluctuations in $y$ are  multifractal.
Table~\ref{tab1}
shows the values of $\alpha_q$ and $\beta_q$ obtained using a fit by
(\ref{mot6}). To avoid trivial effects  due to a
bell-shaped structure of the multiplicity distribution at large $M$,
the  fit is limited to
$\ln M>3$ for $q=2$ and  $\ln M>2$ for $q>2$.

\begin{figure}
\begin{center}

\vspace{-2.5cm}
\mbox{\epsfig{file=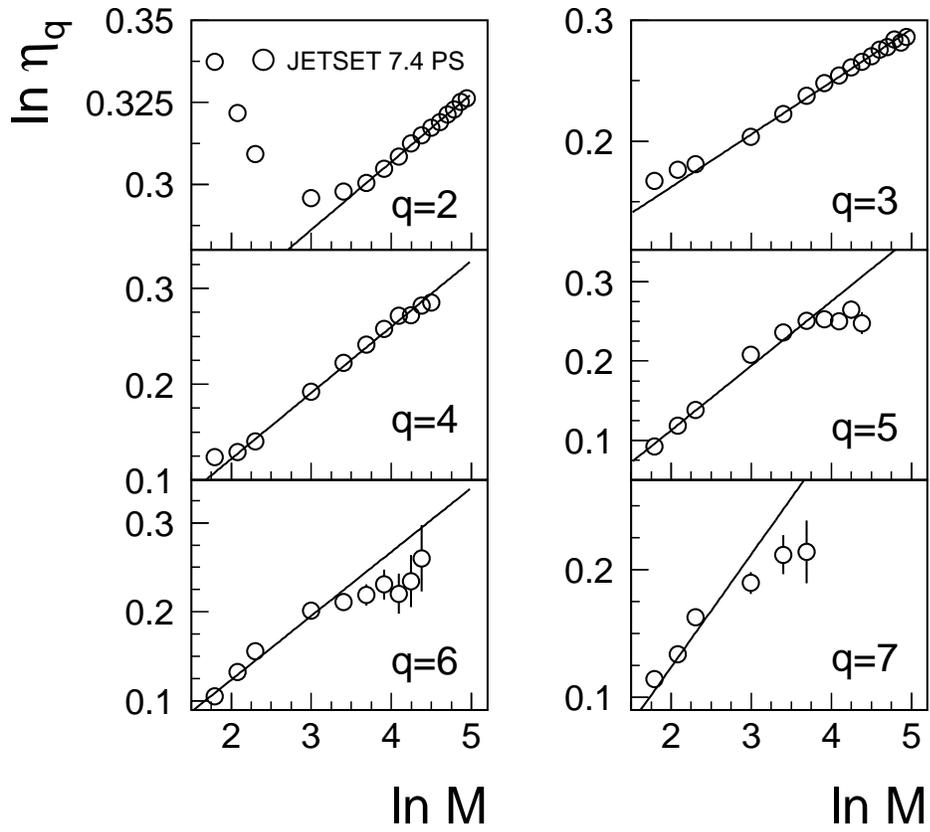,width=13.0cm}}
\caption[bpmark2]
{\it BPs for  rapidity defined with respect
to the thrust axis for JETSET 7.4 PS model.
The lines show the fit by (\ref{mot6}) with the parameters presented in
Table~\ref{tab1}.}
\label{bpmark2}
\end{center}
\end{figure}

Fig.~\ref{bpmark2} and  Table~\ref{tab1} show the predictions
of the JETSET 7.4 PS \cite{jetset} model with the L3 
default parameters \cite{l3def}.
The charged final-state  hadrons  were generated at 91.2 GeV.
The total number of events  is 2.0M.
The regions  $\ln M <3$ (for $\eta_2$ and $\eta_3$) and
$\ln M <2$ (for $\eta_4$) were excluded from the fits.
Note also that $\chi^2$ test for the Monte Carlo is rather poor since,
for the large statistics  used, the behavior of $\eta_q(M)$ shows
a clear complex  structure caused by the
presence of resonance decay products and the
points for different $M$ are not statistically independent.

Table~\ref{tab2} shows the fractal dimensions
$\cd_n$ obtained using (\ref{mul2}). 
The values of $\cd_n$ decrease with increasing $n$,
indicating that the degree of non-homogeneity of the distributions
increases for particles emitted in the cascade later.

%%%%%%%%%%%%%%%%%%%%%%%%%%%%%%%%%%%%%%%%%%%%%%%%%%%%%%%
\section{Model predictions}
%%%%%%%%%%%%%%%%%%%%%%%%%%%%%%%%%%%%%%%%%%%%%%%%%%%%%%% 
\label{mpred}

We have now set up a formalism that  handles  the 
local scale-invariant fluctuations inside a  cascade. 
Qualitatively, the model proposed above 
can reproduce the power-like dependence
of BPs observed in $\E$ data \cite{bpexp} and
other process \cite{bp1}.

A  most direct prediction of this 
approach  is that the power-like behavior 
of the BPs is energy independent: The local fluctuations are determined
by $\g_n(\d )$ in (\ref{18cc}). They, in turn, depend
only on the fractal dimensions 
$\cd_n$. As a result, parameters $\alpha_q$ determining
the phase-space fluctuations in (\ref{mul2}) are $t$-independent. 

The model, however, has only low  predictive  power  unless
we reduce the number of  free parameters $\cd_n$
in (\ref{mul2}). To do this, 
let us rewrite the $\cd_n$ as 
\beq
\cd_n=\cd_0-A_n,
\label{mp1}
\eeq
so that positive $A_n$ represents 
the deviation of fluctuations from
the trivial ones 
($A_n=0$ actually corresponds  to the case of no correlations or
uniform cascade distributions).
We shall call the parameters $A_n$ as the
strength of dynamical correlations
on the $n+1$ multiplicity stage of the branching. 
Since $\cd_{n-1}\geq \cd_n$, we have
\beq
A_0=0\leq A_1\leq A_2\leq \ldots \leq A_n.
\label{asss}
\eeq

The physical meaning of $A_n$ is rather clear:
$A_n$ is determined by the collinear singularities of 
gluon emission and the extent of interference between soft
partons leading to angular ordering.
Generally, however, $A_n$ may absorb many other physical effects
in jet beyond DLLA. This quantity can  incorporate effects from
energy-momentum balance (recoil effect) in two-parton splittings,
heavy quark production and non-perturbative effects: 
hadronization, resonance decay and Bose-Einstein correlations.
Since contributions from these effects are poorly known and
at present cannot be taken into account in analytical calculations,
below we shall make an attempt to treat $A_n$ on a general statistical
ground. 

\medskip

Several remarkable features of $A_n$ are immediately apparent:

a) $A_n$ characterizes a single particle inside $\d\T$
belonging to a system with  $n$  other  particles already produced 
inside this  interval at the previous cascade stages.
    
b) Since $A_n$ is connected with correlations/fluctuations, one can consider
it as a strength of  ``interaction''  of a single particle
with another. According to (\ref{asss}), 
such an interaction becomes stronger with
the increase of  multiplicity $n$. 

\medskip

These two features suggest  that 
$A_n$ is analogous to the binding (pairing)
energy per nucleon in nuclear physics.
Using this analogy, the form of $A_n$ can be 
readily deduced without detailed information  on correlations. 

Let us first consider
the following two extreme  cases:

\medskip

1) Since the Markov chain is based on 
two-particle splittings, one can
assume that there exist positive correlations only between
the particles $a_1$ and $a_2$ of
the two-particle splitting $a_1\to a_1 + a_2$, which is a basic
element of  the  Markov chain. 
From a statistical point of view,
the effect  tends to make two partons
more strongly  bound in  phase space,
i.e., the probability that  particles  $a_1$ and $a_2$ occupy a
very small phase-space bin is larger than
that without dynamical correlations.
After the next splitting of each particle,
one has 2 two-particle pairs.
For an $(n+1)$-particle system, the number
of pairs stemming from  the two-particle splittings is 
$(n+1)/2$, and we can write
\beq
A_n=A^{\S}\frac{n+1}{2},
\label{mp33}
\eeq
where $A^{\S}$ is a constant describing the pair correlation in the case
of  {\em two-particle} correlations.\footnote{The
two-particle and multiparticle correlations
introduced in our statistical model to describe the cascade 
have nothing to do with
the  two-particle and multiparticle correlations
in the final-state hadrons measured by means of the two-particle
and multiparticle correlation functions \cite{int1}. We borrowed  
these terms following an  analogy with the Weizs\"{a}cker mass formula for
the binding energy per nucleon in nuclear physics.}
Note that  the applicability
of (\ref{mp33}) for odd $n$ is only an approximation to 
make the correlations easy to handle. 
We shall correct this expression later. 

\begin{figure}
\begin{center}
\begin{picture}(435,180)
\put(90,0){
         {\epsfig{file=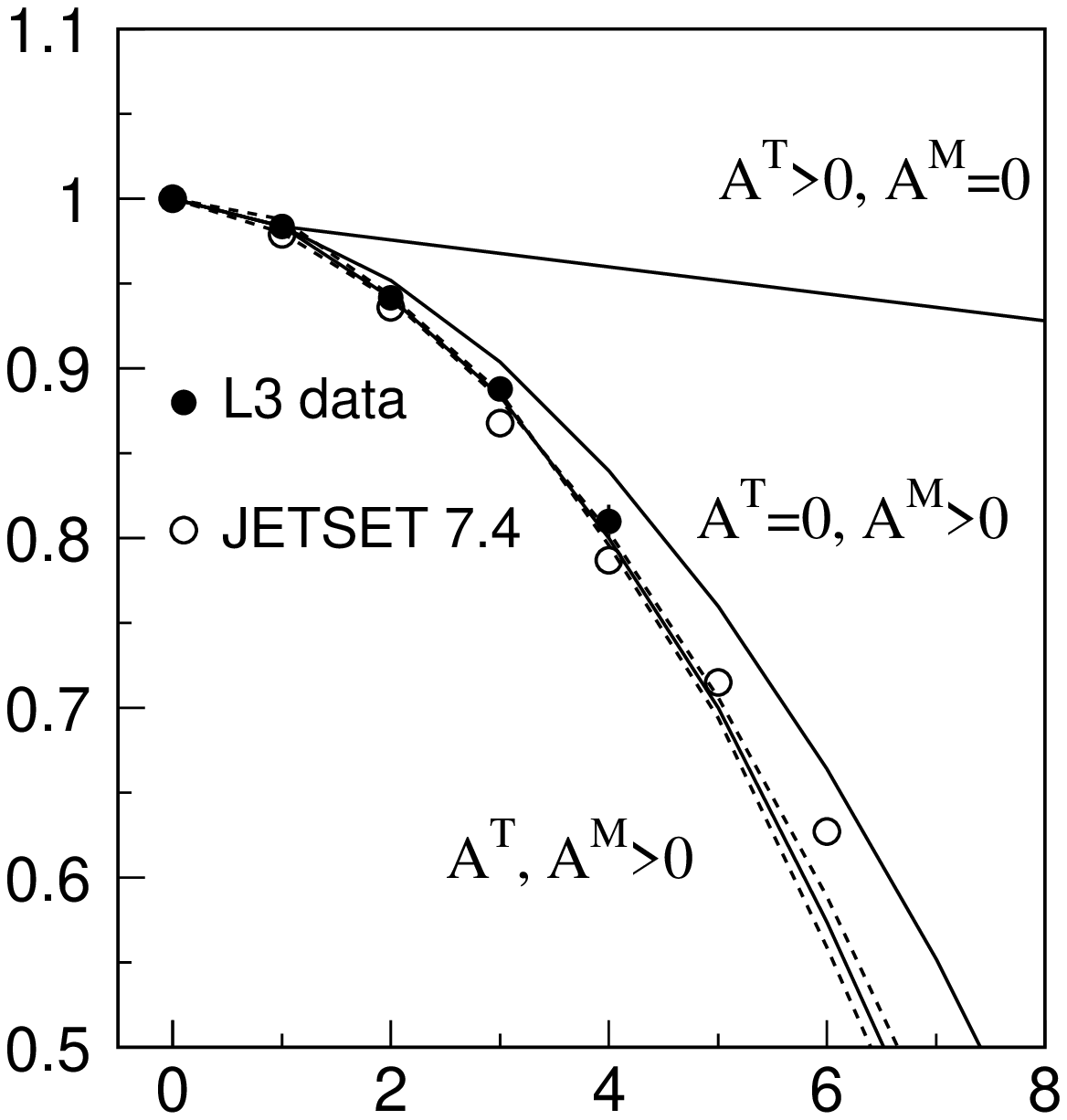, width=8.0cm}}
         }
\Text(310, 5)[c]{\large\boldmath $n$}
\Text(95,195)[c]{\large\boldmath $\mathcal{D}_n$}
\end{picture}
\caption[dim]
{\it The behavior of $\mathcal{D}_n$
for $\E$-annihilation data, JETSET 7.4 PS and
the model predictions for: a) Two-particle
correlations ($A^{\S}>0, A^{\L}=0$);
b) Multiparticle  correlations ($A^{\S}=0$,
$A^{\L}>0$); 
c) Both  two-particle and multiparticle
correlations ($A^{\S}, A^{\L}>0$).}
\label{dimm}
\end{center}
\end{figure}

If  only two-particle correlations  (\ref{mp33}) exist, then one obtains
from (\ref{mp1}) and (\ref{mul2})
\beq
\cd_n = \cd_0 - A^{\S}\frac{n+1}{2},
\label{mp4}
\eeq
\beq
\alpha_2=A^{\S}, \qquad \alpha_{q\ge3}=0.5\, A^{\S}.
\label{mp44}
\eeq
The behavior  $\alpha_q=0.5\alpha_2$  has been 
found to correspond to multiplicity fluctuations  in 
$\mbox{p}\bar{\mbox{p}}$ collisions \cite{bp1}.
However, $\E$  data show a  stronger multifractal signal. 
The behavior of (\ref{mp4}) with $A^{\S}$=0.016  
for the $\E$ data is shown in Fig.~\ref{dimm} ($A^{\S}>0, A^{\L}=0$).
The value of $A^{\S}$ is equal to 
$\alpha_2$ taken from the experimental data (see Table~\ref{tab1}).
The model fails to reproduce the 
$n$-dependence of $\cd_n$ for data and JETSET model.

\begin{figure}
\begin{center}
\begin{sideways}
\begin{sideways}
\begin{sideways}
\mbox{\epsfig{file=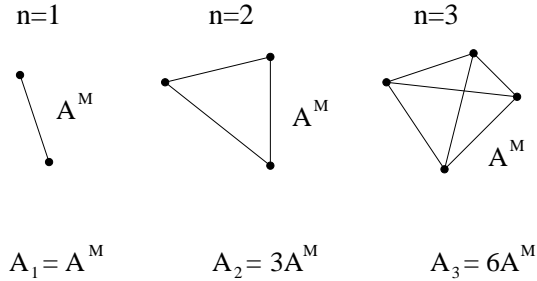,height=7.0cm}}
\end{sideways}
\end{sideways}
\end{sideways}
\caption[sc1]
{\it A schematic representation of the multiparticle
correlations for an  $(n+1)$-particle system ($n=1,2,3$).}
\label{sc1}
\end{center}
\end{figure}

2) Now let us consider another limiting case of correlations.
Let us  assume that each particle of a given $(n+1)$th particle 
generation is attracted in equal extent by all of the  
other  $n$ particles already produced. There are exactly  
$n(n+1)/2$  interactions between  $n+1$ particles
uniformly distributed over the
small phase-space volume. (Such a uniformity must, of course,  be
treated as an  average over all events.)
Hence, the correlation strength is (see Fig.~\ref{sc1})
\beq
A_n=A^{\L}\, \frac{n(n+1)}{2},
\label{mp2}
\eeq
where $A^{\L}$ is a constant characterizing 
the correlation between  any two particles.
It completely determines
{\em many-particle} correlations in such a system.

Having made this simple assumption, one has
\beq
\cd_n= \cd_0 - A^{\L}\frac{n(n+1)}{2}, 
\label{mp3}
\eeq
and, according to (\ref{mul2}), 
the power-law indices for the BPs in the form
\beq
\alpha_2=A^{\L}, \qquad \alpha_{q\ge 3}= A^{\L}(q-1).
\label{mp3x}
\eeq
The result for $A^{\L}=0.016$  is shown  in
Fig.~\ref{dimm} ($A^{\S}=0$, $A^{\L}>0$). As we see, 
this prediction is rather close to the experimental result.
However, it still cannot 
give a satisfactory  description of the  data and  JETSET model.
In fact, such a disagreement is not a surprise since we systematically
ignored the trivial fact that particles 
can interact with different strength.

\medskip

As was mentioned, to some extent,  
$A_n$ is analogous to the binding (pairing)   
energy per nucleon in nuclear physics.
In fact, expression (\ref{mp33}) 
is analogous to the ``volume''
effect  if  the nuclear density is roughly constant.
Then each nucleon  has about the same number of neighbors and
(\ref{mp33}) actually represents the short-range correlations.
Then (\ref{mp2}) 
is analogous to the Coulomb repulsion term in the
Weizs\"{a}cker mass formula which is proportional to
\cite{nucl}
\beq
-\alpha\frac{Z(Z-1)}{2},
\label{mp2w}
\eeq
where $Z$ is the number of protons and
$\alpha =e^2/4\pi$ is the fine-structure constant of QED.
The negative sign implies a reduction in binding energy.
For QCD, of course, the Coulomb interaction
is not the dominant part of the correlations 
and the introduced correlations
should be attributed to other reasons.

Following the same logic, $A_n$ can be constructed
analogously to the semi-empirical Weizs\"{a}cker mass formula by
combining the different types of correlations and taking into
account  the obvious properties of the particle system
in question. To see this, let us consider the following
cascade chain:
$$
a_1\to (a_1+a_2)\to (a_1+a_3) + a_2\to\ldots ,
$$
where the $a_n$ represents  a  parton in 
independent sequential splittings.
The particles in parentheses are pairs arising 
due to two-particle  splitting of  parent particles on each stage. 
It is natural to assume that correlations
between particles in the parentheses   are different from those between
the particles that have already been produced. For example, the particles
in the pairs $(a_1,a_2)$ and  $(a_2,a_3)$ produced 
on the three-particle stage can also be correlated, 
but to an  extent different from 
those in the pair $(a_1,a_3)$  which stem 
directly from the two-particle splitting.
Thus to make a step towards a more realistic description,
it is necessary to take into account a non-homogeneity of
parton interactions in the cascade.

First of all, let us describe the correlations between the particles
in two-particle splittings. 
For this, we should take into account the odd-even effect
in the two-particle correlations which is important for 
small $n$ (this was dropped for simplicity in (\ref{mp33})).  
A corrected expression (\ref{mp33}) reads  as  
\beq
A^{\S}_n\equiv A^{\S}\times\left\{\begin{array}{ll}
(n+1)/2, & \mbox{for $n=1,3,5,\ldots$}\\
n/2,
& \mbox{for $n=2,4,6,\ldots$}
\end{array}\right.
\label{re2}
\eeq

The next step is to take into account the multiparticle
correlations arising  between the
particles  produced in the previous stages
of the cascade. As before, to simplify our considerations, we  assume
that this kind of (multiparticle) correlations can be characterized
by a single parameter $A^{\L}$ responsible for the correlation between
any particles stemming from  {\em  different} parents.
For any $n$-particle system,
the form of these correlations can be obtained
by subtracting  from  a term  of the form  (\ref{mp2}), representing  all
possible pair correlations,  a  term like (\ref{re2}) describing
two-particle  correlations 
which are taken into account by (\ref{re2}). 
The final expression  reads
\beq
A^{\L}_n\equiv A^{\L}\frac{n(n+1)}{2} - A^{\L}\times\left\{\begin{array}{ll}
(n+1)/2, & \mbox{for $n=1,3,5,\ldots$}\\
n/2,
& \mbox{for $n=2,4,6,\ldots$}
\end{array}\right.
\label{re3}
\eeq 

The last step is to combine both  contributions together, 
\beq
\cd_n = \cd_0 - A^{\S}_n - A^{\L}_n,
\label{mp41}
\eeq
\beq
\alpha_2=A^{\S},  \qquad 
\alpha_{q\ge 3}= A^{\S}_{q-1} + A^{\L}_{q-1} - A^{\S}_{q-2} - A^{\L}_{q-2}.
\label{mp42}
\eeq
Expressions (\ref{re2}), (\ref{re3}), (\ref{mp41})
and (\ref{mp42}) explicitly  describe  the  
behavior of the correlations
in the cascade on the basis of the two parameters
$A^{\S}$ and $A^{\L}$. 
The parameter $A^{\S}$ describes the correlation between 
particles stemming from the same parent particle  and
$A^{\L}$ characterizes the correlation between 
the particles coming from  different parents.
As in nuclear physics,\footnote{In nuclear physics
the situation is somewhat different: 
$A_n^{\S}$ provides a ``volume'' binding effect with
positive sign and $A_n^{\L}$ has negative sign that implies
a reduction in binding energy.}
we allow these constants to be adjustable and 
consider $A^{\S}$ and $A^{\L}$ as  free parameters which  can be 
evaluated from the  fit.

The parameters   $A^{\L}$ and $A^{\S}$ can be 
obtained  from the two experimental
parameters  $\alpha_2^{\mbox{exp}}$ and $\alpha_3^{\mbox{exp}}$
describing  the power-law behavior of BPs:
\beq
A^{\S}=\alpha_2^{\mbox{exp}},
\label{mp421}
\eeq
\beq
A^{\L}=\alpha_3^{\mbox{exp}}/2.
\label{mp422}
\eeq
Further evolution of the $\cd_n$ and  
the $\alpha_q$  can be predicted by
the model according to  (\ref{mp41}) and  (\ref{mp42}).
For the $\E$ data presented in Table~\ref{tab1}, one
obtains $A^{\S}=0.016\pm 0.004$ and $A^{\L}=0.021\pm 0.002$.
The predictions for $\cd_n$ 
are shown in Fig.~\ref{dimm}
($A^{\S}, A^{\L}>0$).
The dashed lines show the uncertainty in the behavior of $\cd_n$
due to the statistical errors on  $A^{\S}$ and
$A^{\L}$.
Our predictions  agree with
the experimental data  well.
The agreement with the JETSET becomes better if one uses  the
values of $\alpha_2$ and $\alpha_3$ from the Monte-Carlo model to
determine $A^{\S}$ and $A^{\L}$.

Note that expressions (\ref{mp41}) and (\ref{mp42}) cannot be
valid  for  asymptotically  large $n$ since
the fractal dimensions $\cd_n$ cannot be smaller
then zero.

%%%%%%%%%%%%%%%%%%%%%%%%%%%%%%%%%%%%%%%%%%%%%%
\section{Discussion of the model predictions}
%%%%%%%%%%%%%%%%%%%%%%%%%%%%%%%%%%%%%%%%%%%%%%

One of the striking features of the  results obtained is that
good agreement between the model and  the data  
is possible only if the value of 
$A^{\S}$  is smaller than that of $A^{\L}$.
This means that  the binding effect between  two particles
from the same parent  must  be  
smaller than that between  particles
produced earlier from different parent particles, i.e., 
the particles originating from   different parents 
have a larger chance of being  emitted very close to each other.   

There are a number of possible explanations for  this effect.
If one believes that the model  describes the perturbative QCD cascade, 
the reason for this 
may come directly from the color coherence effects. Indeed,
the fact that $A^{\L}> 0$ can be due to the angular ordering:
For a given cascade stage with multiplicity $n$,  collective
correlation effects should exist between each particle
due to the angular ordering history of the previous stages.
Then the smallness of $A^{\S}$ can be explained    
by recoil effects  and the minimal value of the relative transverse
momentum $k_{\bot}$ of decay products in the cascade evolution,
in order to ensure that
partons have enough time to radiate, in their turn, new offspring \cite{bas}.
The latter effect  leads to a  restriction on  the relative
emission angels between the particles $a_1$ and $a_2$  in
the two-particle splitting $a_1\to a_1 + a_2$.
From a statistical point of view,
the effect  tends to make the two partons
less tightly bound in  phase space,
i.e., the probability that both $a_1$ and $a_2$ particles occupy a
very small phase-space bin  is less  than
that without the restriction on the  angle.
If the reason for the condition $A^{\S}<A^{\L}$ indeed  comes from
perturbative QCD, $A_n^{\S}$ has to be connected with the 
momentum transfer cut-off $Q_0$ that limits the  relative
$k_{\bot}$ and plays the role of an effective mass of a parton.

On the other hand, it is reasonable to think that the
proposed formulation of the branching process is sufficiently general
and  can utilize non-perturbative effects as well. 
In fact, the branching 
can be attributed to a certain degree to
hadronization and resonance decay.
Then,  the multiparticle correlations
can arise due to the color exchange between the partons at
the end of the  perturbative regime of QCD branching, necessary for
parton discoloration. Furthermore, if the partons
are replaced by hadrons,  the  large multiparticle correlations
can be attributed to Bose-Einstein interference  between identical pions,
since  these particles 
are usually  produced by different parent ones. 
Then  the smallness  of  $A_n^{\S}$ can be explained  
by an anti-correlation trend between  decay products of resonances.

Note also that  the model can be used for various  complex non-point-like
processes. In this context, one can consider the evolution
of the multiplicity distributions for clusters,
fireballs, resonances {\em etc.},
taking into account peculiar features of these processes and introducing
additional (or other) correlation terms in (\ref{mp41}).

%%%%%%%%%%%%%%%%%%%%%%%%%%%%%%%%%%%%%%%%%%%%%%%%%%%%%%%
\section{Summary and conclusion}
%%%%%%%%%%%%%%%%%%%%%%%%%%%%%%%%%%%%%%%%%%%%%%%%%%%%%%% 

In this paper we developed a new  concept of 
local scale-invariant fluctuations  in branching processes.
In contrast to the approaches  based on   many-particle QCD
correlation  functions \cite{an1,an2} 
and phenomenological continuous densities \cite{rcas}, 
we adopted  a  method based on single-particle probabilities
(or single-particle probability densities) for  each cascade  stage.
They are characterized by the fractal dimensions $\cd_n$
determining a non-uniformity in phase-space distributions for
each particle emitted into a small phase-space domain. 
Such an   idea simplifies the 
picture of phase-space organization of particles inside 
the cascade and allows us  to take into account
the finiteness of the number of particles in the cascade (or finite 
energy), QCD color coherence effects 
and a heterogeneity of correlations between partons
belonging to the different cascade generations.  

The fractal dimensions $\cd_n$
can be experimentally observed by calculating the BPs
which resolve the difference $\cd_{n-1}-\cd_{n}$, according to 
(\ref{mul2}).
A less direct way to measure $\cd_n$ can be performed
from the study of the normalized factorial moments (see (\ref{fmop9})).

The model suggests and makes experimentally
accessible  new physical quantities -  pair correlation
coefficients $A^{\L}$ and $A^{\S}$ determining $\cd_n$. 
The fact that none of these parameters are  zero is due to the 
collinear singularities of the emission probabilities of soft partons. 
However, the way how these parameters determine the directly
observable $\cd_n$ can be due to many reasons. 
In this paper we suggest such a relationship
using  a general  statistical formalism, which,  
in terms of QCD, may absorb the details 
of coherence effects, high-order  
perturbative corrections, recoil effects 
and non-perturbative phenomena, i.e. all the effects
which at present can be  combined together only on the basis of
Monte-Carlo simulations.  
We allow $A^{\L}$ and $A^{\S}$ to be adjustable
that ultimately leads to good quantitative agreement 
with the local fluctuations in $\E$ processes.

The model predicts that the 
experimentally observable parameters $\cd_n$ determining 
the scale-invariant behavior
of BPs $\eta_q(\d )$ are  energy independent. In addition,
they do not depend on details of Markov equation in  
the full phase space.
Both  features  follow from the factorization scheme used to derive
the local fluctuations from the classical Markov branching equation  
for  jet  evolution and the angular ordering scheme
which helps to construct the local version of this equation.
Therefore, to check this approach, precise data on the behavior
of the BPs with energy are needed.

Another model prediction is a suppression of 
positive correlations between the
off-spring particles, $A^{\L}>A^{\S}$, a feature which can 
directly be detected from the study of $q$-dependence of the BPs.
This prediction is also model dependent and the next step
would be to understand how this  effect can be changed  if one
uses  another  physical motivated parameterizations.

In spite of its simplicity, the model describes the correlations 
between partons in branchings  beyond the scope of the Leading
Log Approximation of QCD.  To   leading order in $\ln Q^2$, partons
are free elementary quanta. Evidently, this situation corresponds
to the particular case $\cd_n=1$ (for all $n$) in our scheme. 
Since the model is constructed on the basis of angular ordering,
it takes  advantage of the DLLA. 
However, for very small $\d$, the perturbative QCD ceases to be valid,
since $Q_0$ sets the limit of validity of the smallest bin size and 
perturbative expansion of  QCD. Hence, dealing with 
very small phase-space intervals, our model goes beyond the
perturbative QCD approximations 
studied in \cite{an1}. At the same time, the model can
take into account non-perturbative effects which are  important
if one goes beyond single-particle densities. 
It is evident that the price to pay for this
progress in the description of multiparticle  
correlations inside jet is a purely statistical formalism
eliminating the momentum dependence.

\bigskip
Acknowledgments
\medskip

This work was started during my stay at the High Energy Physics
Institute Nijmegen (HEFIN, The Netherlands).
I thank  W.Kittel 
for reading a first preliminary draft of this  manuscript and
for suggesting improvements.

{}

%%%%%%%%%%%%%%%%%%%%%%%%%%%%%%%%%%%%%%%%%%%%%%%%%%%%%%%%%%%%%%%%%%%

\newpage

\begin{table}[hhh]
\begin{center}
$$\begin{array}{|l|c|c|c|c|c|c|} \hline
  & \alpha_q & \beta_q & \chi^2/$df$ & \alpha_q &
  \beta_q & \chi^2/$df$\\ \hline\hline
& \multicolumn{3}{c|}{$data$} &  \multicolumn{3}{c|}{$JETSET 7.4 PS$}
\\ \hline
q=2     &  0.016 \pm 0.004     & 0.244 \pm 0.018  &  0.8/8  &
0.0206 \pm 0.0005 & 0.224 \pm 0.002 & 2.4/11 \\
q=3     &  0.042 \pm 0.003   & 0.08 \pm 0.01   &  8/12 &
0.0434 \pm 0.0007 & 0.075 \pm 0.003 & 22/13 \\
q=4     &  0.062 \pm 0.004   & 0.01 \pm 0.01   &  9/12 &
0.068 \pm 0.001 &  -0.016 \pm 0.004 & 36/10 \\
q=5     &  0.071 \pm 0.008   & -0.03 \pm 0.02   &  14/11 &
0.081 \pm 0.002 & -0.049 \pm 0.004 & 91/10 \\
q=6     &  -        & - & - &
0.072 \pm 0.002 & -0.019 \pm 0.005 & 48/10 \\
q=7     &  -        & -  & - &
0.088 \pm 0.003 & -0.053 \pm 0.006 & 64/8 \\
\hline
\end{array}$$
\caption[tab1]
{\it Fit results for $\eta_q(M)$ obtained from
the $\E$  data \cite{bpexp}.
The linear function (\ref{mot6}) is used.}
\label{tab1}
\end{center}
%  \end{table}

\vspace{1.0cm}

\vspace{1.0cm}
%  \begin{table}[hhh]
\begin{center}
$$\begin{array}{|l|c|c|} \hline
   & $data$ & $JETSET 7.4 PS$
\\ \hline\hline
n=0     &  1.0      & 1.0 \\
n=1     &  0.984 \pm 0.004   & 0.9794 \pm 0.0005 \\
n=2     &  0.942 \pm 0.005   & 0.936 \pm 0.001 \\
n=3     &  0.888 \pm 0.006   & 0.868 \pm 0.001 \\
n=4     &  0.81  \pm 0.01    & 0.787 \pm 0.002 \\
n=5     &   -                & 0.715 \pm 0.003 \\
n=6     &   -                & 0.627 \pm 0.004 \\
\hline
\end{array}$$
\caption[tab2]
{\it
The values of fractal dimensions $\cd_n$ obtained from
the experimental data and JETSET 7.4 PS.
(see (\ref{mul2a}) and Table~\ref{tab1}).}
\label{tab2}
\end{center}
\end{table}

\newpage

%%%%%%%%%%%%%%%%%%%%%%% FIGURE 1 %%%%%%%%%%%%%%%%%%%%%%%%
%%%%%%%%%%%%%%%%%%%%%%%%%%%%%%%%%%%%%%%%%%%%%%%%%%%%%%%%%

\newpage
%%%%%%%%%%%%%%%%%%%% FIGURE 2 %%%%%%%%%%%%%%%%%%%%%%%%%%%%%%
%%%%%%%%%%%%%%%%%%%%%%%%%%%%%%%%%%%%%%%%%%%%%%%%%%%%%%%%%%%%%

\newpage
%%%%%%%%%%%%%%%%%%%% FIGURE 3 %%%%%%%%%%%%%%%%%%%%%%%%%%%%%%
%%%%%%%%%%%%%%%%%%%%%%%%%%%%%%%%%%%%%%%%%%%%%%%%%%%%%%%%%%%%%

%%%%%%%%%%%%%%%%% FIGURE 4 %%%%%%%%%%%%%%%%%%%%%%%%%%%%%%%%%
\newpage

\bigskip

%%%%%%%%%%%%%%%%%%%%%%% FIGURE 5 %%%%%%%%%%%%%%%%%%%%%%%%
%%%%%%%%%%%%%%%%%%%%%%%%%%%%%%%%%%%%%%%%%%%%%%%%%%%%%%%%%

\end{document}